\documentclass[lettersize,journal]{IEEEtran}
\usepackage[colorlinks,urlcolor=blue]{hyperref}
\usepackage{amsmath,amsfonts}
\usepackage{algorithmic}
\usepackage{algorithm}
\usepackage{array}
\usepackage[caption=false,font=normalsize,labelfont=sf,textfont=sf]{subfig}
\usepackage{textcomp}
\usepackage{subfig}
\usepackage{stfloats}
\usepackage{url}
\usepackage{verbatim}
\usepackage{bbm}
\usepackage{caption}
\usepackage{graphicx}
\usepackage{multirow}
\usepackage{multicol}
\usepackage{cite}
\usepackage{capt-of,etoolbox}

\begin{document}

\title{UniHM: Universal Human Motion Generation with Object Interactions in Indoor Scenes}


\author{Zichen~Geng,
        Zeeshan~Hayder,
        Wei~Liu,
        and~Ajmal~Mian,~\IEEEmembership{Senior Member,~IEEE}%
\thanks{This work is supported by UWA ARC 3D Computer Vision HDR Scholarship.}
\thanks{Zichen Geng, Wei Liu, and Ajmal Mian are with the School of Electrical, Electronic and Computer Engineering, University of Western Australia, Perth, WA 6009 Australia (e-mails: zen.geng@research.uwa.edu.au; wei.liu@uwa.edu.au;ajmal.mian@uwa.edu.au).}%
\thanks{Zeeshan Hayder is with the Commonwealth Scientific and Industrial Research Organisation (CSIRO), Canberra, ACT 2601, Australia (e-mail: zeeshan.hayder@data61.csiro.au).}%
}

\markboth{IEEE TRANSACTIONS ON VISUALIZATION AND COMPUTER GRAPHICS}%
{Shell \MakeLowercase{\textit{et al.}}: A Sample Article Using IEEEtran.cls for IEEE Journals}

\twocolumn[{%
\renewcommand\twocolumn[1][]{#1}%
\maketitle
\begin{center}
    \centering
    \captionsetup{type=figure}
    \includegraphics[width=\textwidth]{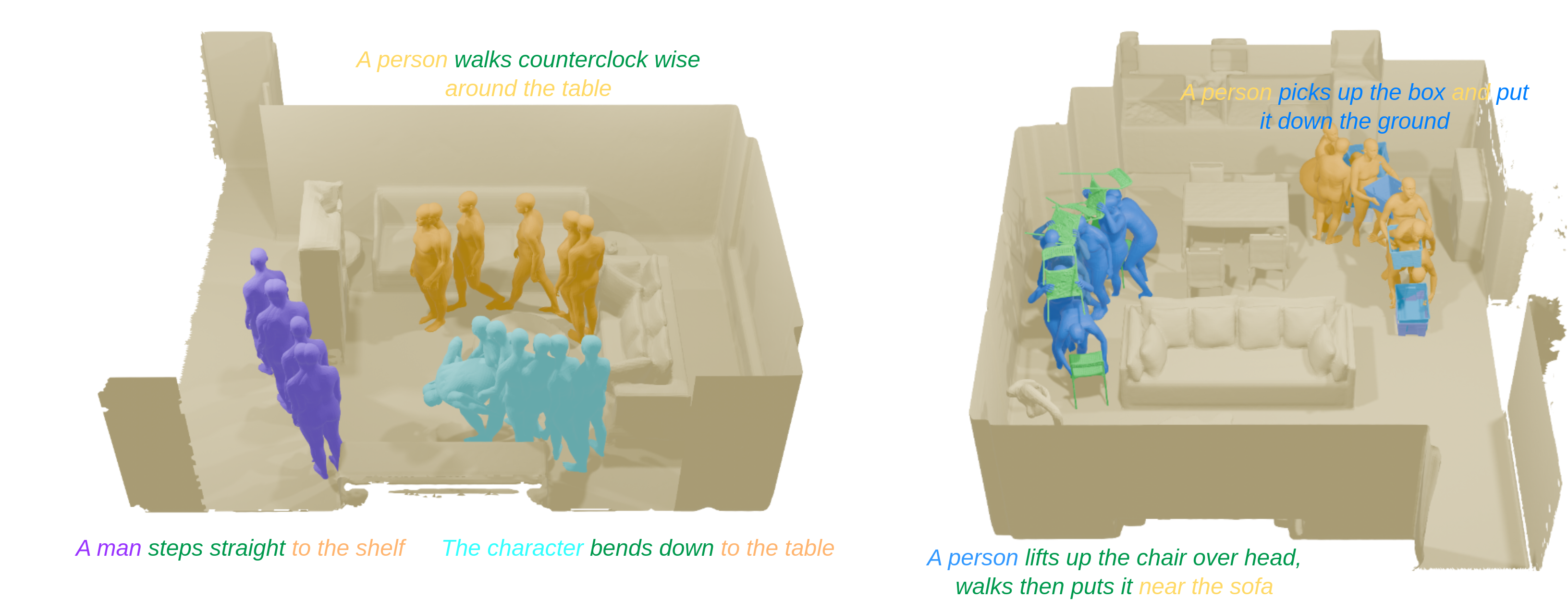}
    \captionof{figure}{Text-to-Motion sequences (left) and Text-to-HOI sequences (right) generated by our approach.}
\end{center}%
}]

\begin{abstract}
Human motion synthesis in complex scenes presents a fundamental challenge, extending beyond conventional Text-to-Motion tasks by requiring the integration of diverse modalities such as static environments, movable objects, natural language prompts, and spatial waypoints. Existing language-conditioned motion models often struggle with scene-aware motion generation due to limitations in motion tokenization, which leads to information loss and fails to capture the continuous, context-dependent nature of 3D human movement. To address these issues, we propose UniHM, a unified motion language model that leverages diffusion-based generation for synthesizing scene-aware human motion. UniHM is the first framework to support both Text-to-Motion and Text-to-Human-Object Interaction (HOI) in complex 3D scenes. Our approach introduces three key contributions: (1) a mixed-motion representation that fuses continuous 6DoF motion with discrete local motion tokens to improve motion realism; (2) a novel Look-Up-Free Quantization VAE (LFQ-VAE) that surpasses traditional VQ-VAEs in both reconstruction accuracy and generative performance; and (3) an enriched version of the Lingo dataset augmented with HumanML3D annotations, providing stronger supervision for scene-specific motion learning. Experimental results demonstrate that UniHM achieves comparative performance on the OMOMO benchmark for text-to-HOI synthesis and yields competitive results on HumanML3D for general text-conditioned motion generation.

\end{abstract}

\section{Introduction} \label{sec:intro}

Human motion synthesis in complex scenes represents a challenging extension of the Text-to-Motion paradigm, with potential applications in virtual reality, robotics, and interactive environments where accurately synthesized human motion is critical to user experience. While language models have demonstrated considerable success in generating realistic human motion sequences based on text prompts, they struggle to achieve similar efficacy in scene-specific motion generation. Scene-based human motion synthesis requires not only an understanding of human motion but also an intricate integration of diverse modalities, such as static scene elements, moveable objects, text prompts, and motion waypoints. These modalities add layers of complexity that go beyond standard Text-to-Motion tasks, demanding a cohesive synthesis of environmental context and dynamic interaction. In this paper, we are more interested in finding a relatively suitable place for a motion to take instead of actual human scene interaction.

One significant challenge in applying language models to motion generation lies in the limitations of tokenization. Motion tokenization has proven effective in motion synthesis tasks by compressing motion sequences into discrete tokens at small intervals. However, human motion, which involves six degrees of freedom (6DoF), cannot be accurately represented through a globally discrete token system. Human movement is inherently continuous and complex, encompassing transformations in three-dimensional space where movements are highly context-dependent and interrelated. For instance, the same motion sequence can be performed with different orientations and starting positions. In scene-aware motion generation, trajectories and orientations must dynamically adapt to both static and movable scene elements, making discrete tokenization an inadequate representation. This process inevitably results in information loss, reducing the fluidity and realism of the generated motion. To address this issue, we propose a mixed-motion representation that combines continuous 6DoF with discrete local motion tokens.

Another challenge in leveraging language models for indoor motion generation is the presence of high-frequency micro-movements during interactions. These subtle motions create a high density of similar motion clips in the quantization space, leading to poor generation quality due to overlapping or redundant representations. This issue has been extensively studied in the context of image generation using discrete tokens. In image generation, it is well established that a larger vocabulary size combined with a smaller embedding dimension improves generation quality \cite{lfqvae}. Inspired by these insights, we explore strategies to enhance tokenization for motion generation, ensuring higher fidelity and diversity in synthesized motions.
Further complicating the adaptation of language models to scene-based human motion synthesis is the challenge of defining an appropriate loss function. While traditional loss functions, such as cross-entropy, are effective for tasks involving discrete outputs, they fail to evaluate the intricate requirements of continuous motion within a complex scene. To this end, we propose a mixed diffusion model that predicts an initial next motion token while bridging the 6DoF and discrete motion tokens. This approach aims to better capture the contextual appropriateness of motion in relation to static elements, moveable objects, and human-object interactions, which are not adequately addressed by conventional loss metrics.
Therefore, while autoregressive and token-based models have advanced Text-to-Motion tasks, new methods are essential for scene-based human motion synthesis. Effective solutions must handle continuous motion trajectories without sacrificing spatial realism or contextual coherence. Developing approaches that can navigate the complexities of scene modalities—including human motion, static and moveable objects, and text prompts—will be crucial to advancing realistic scene-based motion synthesis. Furthermore, data availability remains a limiting factor; to address this, we augment the existing datasets, such as the Trumans Dataset, with HumanML3D annotations, enriching the data for better scene-specific motion learning.
Our contributions can be summarized as follows:
\begin{itemize}
    \item Foundation Model: We propose a foundational model UniHM for Human Motion Synthesis in scenes based on a motion language model integrated with a diffusion model. Our UniHM can take text, waypoints, and scenes as conditions in a single model. This architecture enables greater flexibility and expandability for synthesizing motion within complex scenes.
    \item Tokenization for 3D Sequences: We address the limitations of tokenization in representing 3D motion sequences by developing a mixed-motion representation that combines continuous 6DoF with discrete local motion tokens.
    \item Look-Up-Free Quantization VAE (LFQ-VAE): We introduce a Look-Up-Free Quantization Variational Autoencoder (LFQ-VAE), which surpasses the traditional Vector-Quantized VAE (VQ-VAE) in both reconstruction fidelity and generative capacity.
    \item Dataset Augmentation: To enrich data quality and variety, we augment the Lingo Dataset with the HumanML3D dataset, populating it with motion sequences in scenes and adding extensive annotations for enhanced scene-specific training.
\end{itemize} 
These contributions collectively address the existing limitations in scene-based human motion synthesis, pushing the boundaries of Text-to-Motion modeling, and enabling richer, more contextually accurate motion generation for practical applications. Our approach achieves comparable results on the HumanML3D dataset for the Text-to-Motion task, and the OMOMO dataset for the Text-to-HOI task.

\section{Related Work}
\label{sec:related}

\subsection{Text-to-Motion}
The field of Text-to-Motion generation has gained significant traction with the emergence of diffusion models \cite{ddpmo} and the establishment of foundational benchmarks \cite{a2m}. Early methods predominantly relied on VAE-based approaches \cite{a2m, temos} and GANs \cite{xu2023actformer}, but recent advancements have led to two dominant paradigms: diffusion-based and autoregressive models.

Diffusion models were first adapted for Text-to-Motion by \cite{mdm, mofusion, motiondiffuse, kim2022flame}, leveraging transformers to directly model raw motion sequences. Later, \cite{mld} introduced the first latent-space diffusion model, incorporating a Transformer-based VAE for improved motion quality and efficiency.

In contrast, autoregressive approaches model motion as discrete sequences. \cite{t2mgpt} pioneered the use of a CNN-based VQ-VAE in conjunction with GPT to generate motion token-by-token. Subsequent research \cite{motiongpt, guo2024momask, attt2m} refined this vector quantization strategy, enhancing the autoregressive framework. Some hybrid models combine both paradigms, such as cross-attention denoisers \cite{armd} and MLP-based autoregressive motion predictors \cite{cmdm}. However, conventional autoregressive models primarily condition predictions on previous frames, limiting their ability to capture long-term context and making them susceptible to error accumulation over time.

To address these limitations, \cite{ardiff} introduced a new paradigm that bypasses vector quantization, instead predicting motion token distributions in continuous space using an autoregressive diffusion framework. This method effectively retains contextual consistency across sequences, marking a significant advancement in Text-to-Motion generation.

\subsection{Text-to-HOI}
Compared to Text-to-Motion, Text-to-HOI generation remains a relatively underexplored domain. Early research focused on generating human-object interactions in static environments \cite{humanise, GTAIM, playforbenchmark, PROX} or pre-defined HOI actions \cite{posa, MIME}. Additionally, reinforcement learning approaches have been explored to synthesize physically plausible human motion in interactive settings \cite{physhoi}.

Recently, diffusion models have been adopted for HOI generation. \cite{interdiff} introduced the first autoregressive diffusion model for HOI, though it primarily conditions on historical motion sequences rather than text. The emergence of text-conditioned HOI models \cite{cghoi, hoidiff} has further advanced the field. \cite{cghoi} proposed a method that guides diffusion using contact data, ensuring realistic object interaction at each time step. Similarly, \cite{hoidiff} emphasized the importance of affordance constraints, integrating an affordance-generation module and applying post-optimization after the inverse diffusion process.
Building upon these foundations, \cite{thor} introduced a relation-intervention diffusion model, refining object rotation dynamics throughout the inverse diffusion process. On a different trajectory, \cite{chois} proposed an autoregressive HOI model, leveraging text, object mesh, and initial states to predict future interactions. More recently, \cite{interdreamer} introduced InterDreamer, a World Model-based approach that combines LLM-generated textual planning with pre-trained Text-to-Motion models, enabling zero-shot HOI synthesis. These advancements highlight the growing potential of HOI generation in dynamic and interactive environments.

\subsection{Human Scene Interaction}

Human-Scene Interaction (HSI) is a broad yet underdeveloped research field compared to other motion generation tasks, primarily due to the complexity of scene representations and the diverse modalities involved. Existing approaches can be categorized into two main groups: data-driven methods and physics-based methods.
Data-driven approaches leverage real-world motion data to synthesize human-scene interactions. Early pioneers, such as PROX \cite{PROX} and HUMANISE \cite{humanise}, incorporated human motion with scene point clouds, utilizing models like PointNet++ \cite{pointnet++} and Point Transformer \cite{pointtrm} to encode spatial constraints and guide motion synthesis. More recent works, such as TRUMAN and LINGO \cite{trumans, lingo}, introduced voxel-based scene representations with text annotations, improving scene-aware motion generation through robust autoregressive diffusion models and ego-centered mesh encoding for contextual understanding.
In contrast, physics-based methods focus on physically plausible motion synthesis by integrating simulation environments. For instance, SceneDiff \cite{scenediff} and UniHSI \cite{unihsi} utilize IsaacSim \cite{isaac} for reinforcement learning-based motion generation, ensuring realistic scene interactions. However, while these methods enforce physical feasibility, they often struggle to preserve natural human motion fidelity compared to data-driven approaches.
In this work, we focus on scene-aware motion generation, particularly in collision avoidance and navigation, to enhance the realism and practicality of human-scene interactions.

\section{Method}\label{sec:method}

\begin{figure*}
    \centering
    \includegraphics[width=\linewidth]{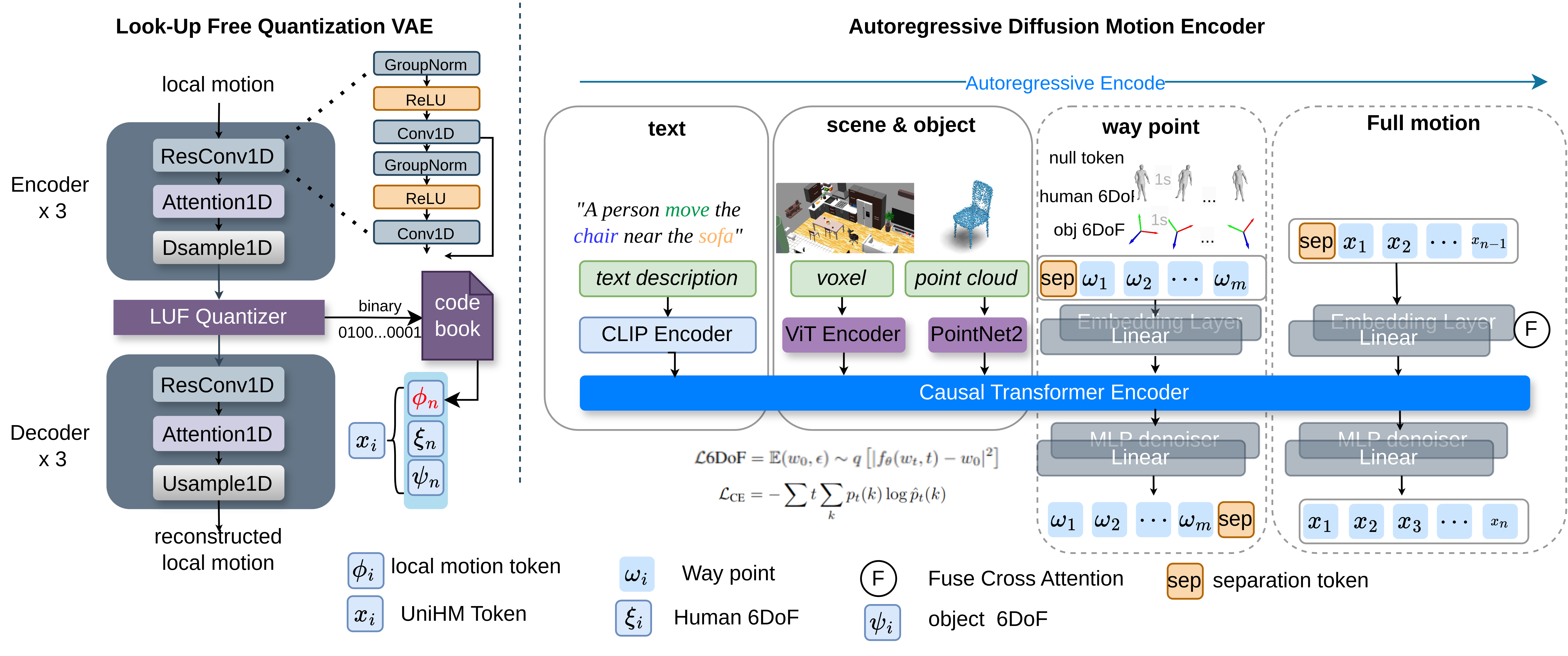}
    \caption{UniHM architecture and Look-up Free Quantization Variational Autoencoder.}
    \label{fig:architecture}
\end{figure*}

UniHM shown in Figure \ref{fig:architecture} is a novel framework for human motion generation with object interactions, leveraging autoregressive text, scene voxel, and object point cloud inputs. The proposed method predicts sparse waypoints representing the 6-DoF motion of humans and objects over a 1-second window and subsequently refines these waypoints into a full motion sequence using a large language backbone. The motion generation pipeline consists of two primary stages: (1) discrete autoregressive generation of local-motion tokens via a Contextual Motion Embedding (CME), and (2) a lightweight MLP-based denoiser that samples 6-DoF trajectories from the CME. 

\subsection{UniHM Dataset}
 To combine multiple tasks in one dataset. We propose a synthesized dataset called \textbf{UniHM} by populating the largest Text-to-Motion dataset, HumanML3D dataset, and the largest Text-to-HOI dataset, OMOMO dataset together with a newly proposed Human-Scene-Interaction dataset, Lingo dataset with text annotations. For a uniform representation, we modify the HumanML3D dataset from a canonical representation to a global SMPL representation. Our UniHM dataset consists of 44962 sequences with a maximum length of 300 frames at 30 fps. Each sequence contains a motion sequence, 1-3 text descriptions, a scene voxel, and a point cloud of randomly sampled 1024 points. 
 
 To populate the motion sequences in HumanML3D into the scene, we first remove the scene relative sequences for example sitting on an invisible chair. Then, the motion sequences will be placed in the center of the unoccupied voxel and a translation and rotation offset will be optimized to give a minimum collision score. For the motion sequences that are not capable of populating in scenes, they will be regarded as scene-less motion sequences and a null scene will be given for modality consistency. 
 
 To provide more scene-aware descriptions, we first render the scene and corresponding human \& object motions to video and then utilize VideoLlama2 to generate new scene-aware descriptions based on the original text descriptions.

The data representation is organized as follows:
\noindent \textbf{Motion Representation}: We utilize the SMPL representation $\mathcal{M} \in \mathcal{R}^{T,69}$ with 22 joints, whose data representation consists of global root translations, orientations, and local joint rotations in axis angle. To cope with the object motion, an additional object 6DoF is concatenated with the SMPL motion representation as $\mathcal{M} \in \mathbb{R}^{T, 75}$ 

\noindent \textbf{Text Prompt:} A natural language description guiding the intended motion behavior. Following previous work, we use the CLIP model to encode the text feature to condition the motion sequences. 

\noindent \textbf{Scene Representation:} The scene in this task, unlike previous work using point cloud, is represented in a voxel $\mathcal{S} \in \{0,1\}^{H, W, D}$, where $H, W$ are the length and width of the XZ plane of the scene, $D$ is the height of the Y axis and 1 for occupied scene and 0 for not. Such a voxelized encoding of the environment provides spatial constraints. We train a ViT-based \cite{vit} VAE with KL diversity to learn compressed scene embeddings. We randomly replace corresponding scenes with a null scene and use canonical motion representation to enhance the motion understanding for text-to-motion and text-to-HOI tasks.

\noindent \textbf{Object Representation} is essential for the text-to-HOI task. We populate the OMOMO dataset into the scene 3D-Loco Front dataset. The point cloud is encoded by a PointNet++ trained for a segmentation task. An empty object representation token will be given for those motion sequences without objects to interact with.

\subsection{Look-Up Free Quantization VAE}
To enhance motion tokenization, UniHM employs a {Look-Up Free Quantized Variational Autoencoder (LFQ-VAE)}.
Traditional vector quantization approaches like VQ-VAE rely on a high dimensional parameterized codebook, e.g. 512 \cite{t2mgpt}. In contrast, LFQ-VAE eliminates the explicit codebook by reducing its dimensionality to zero, replacing the codebook $\mathbf{C} \in \mathcal{R}^{K\times d}$ by $\mathbb{C}=K$. This improves generalization, prevents overfitting to predefined motion patterns, and allows for more expressive motion representations. Besides, a larger vocabulary size can learn high-frequent micro-motion such as small clapping, and small nods. 

Among multiple quantization methods, we applied the most straightforward one following the work, which utilizes independent codebook dimensions and binary latent. The latent space of the Look-up Free Quantization is decomposed as the Cartesian product of single-dimension variables, as $\mathbb{C} = \times_{i=1}^{\log_2K} C_i$. For a given feature vector $\mathbf{z} \in \mathcal{R}^{\log_2K}$, each dimension of the quantized representations $q(\mathbf{z})$ is obtained by:
\begin{equation*}
q(z_i) = C_{i,j}, \text{where} j = \arg \min_k ||z_i - C_{i,j}||,
\end{equation*}

where $C_{i,j}$ is the $j$-th value in $C_i = \{1, -1\}$.

\noindent The token index for $q(\mathbf{z})$ is given by:

\begin{align}
    \text{Index}(\mathbf{z}) &= \sum_{i=1}^{\log_2K} \arg \min_k ||z_i - C_{i,k}||\prod_{b=0}^{i-1}|C_b|  \\
    & = \sum_{i=1}^{\log_2K} 2^{i-1} \mathbbm{1} \{z_i > 0\}
\end{align}

The encoder and decoder utilize a similar structure, which consists of a 1-D CNN block with residual connections, followed by an attention layer and a CNN downsample/upsample block. 
The encoder maps motion sequences into a latent space where motion tokens are continuously learned without discrete lookup tables. This ensures better adaptability to unseen motions. Instead of selecting nearest neighbor tokens from a pre-defined codebook, our approach directly parameterizes motion via binary representation, optimizing reconstruction quality while maintaining diversity. Compared with VQ=VAE, LFQ-VAE ensures smooth transitions between motion states by increasing vocabulary size and reducing quantization artifacts that cause high proxy in discrete space. To further increase the utilization of the codebook, following the implementation of \cite{lfqvae}, we add an entropy loss to encourage a more diverse 

\begin{equation*}
    \mathcal{L}_{\text{entropy}} = \mathbb{E}[H(q(\mathbf{z}))] - H [\mathbb{E}(q(\mathbf{z}))],
\end{equation*}
where $H$ is the Shannon entropy of the latent distribution. The total loss function could be generated by:

\begin{equation}
    \mathcal{L}_{\text{lfq}} = \lambda_{\text{recon}} \mathcal{L}_{\text{recon}} 
    + \lambda_{\text{commit}} \mathcal{L}_{\text{commit}} 
    + \lambda_{\text{entropy}} \mathcal{L}_{\text{entropy}} 
\end{equation}

where $\lambda_{\text{recon}}, \lambda_{\text{commit}}, \lambda_{\text{entropy}}, \lambda_{\text{KL}}$ are weighting coefficients. The reconstruction loss ensures accurate motion recovery:

\begin{equation}
    \mathcal{L}_{\text{recon}} = \mathbb{E}_{x \sim D} \left[ || x - \hat{x} ||_2^2 \right]
\end{equation}

where $x$ is the original motion sequence, and $\hat{x}$ is the reconstructed sequence. The commitment loss prevents latent vectors from fluctuating excessively:

\begin{equation}
    \mathcal{L}_{\text{commit}} = || \text{sg}[z] - \bar{z} ||_2^2
\end{equation}

where $\text{sg}[\cdot]$ is the stop-gradient operator, $z$ is the latent representation, and $\bar{z}$ is the quantized latent vector.

\subsection{Sparse Waypoint Generation}
Given the multimodal input, UniHM first generates sparse waypoints representing continuous 6-DoF poses of the human and interacting objects. We follow the setting of CHOIS and pick the waypoint for every 1 second. Since the waypoint is a sparse signal, we repeat them for a number of the token frames (8). Since the autoregressive backbone needs to process both the 6-DoF and motion tokens in the later full-motion generation, modality types must be consistent when encoding initial waypoints. Therefore, we add an ad-hoc null motion token with each waypoint to make the input modality consistent with the subsequential full-motion sequences. 

This stage is handled by an autoregressive module that predicts waypoints conditioned on past waypoint history and contextual embeddings: The waypoints will be projected through a linear layer and the null tokens will be encoded by an embedding layer. A cross-attention layer will serve as a feature fusion module to fuse the waypoint feature and the motion token embeddings. The fused feature will be fed into a causal Transformer encoder, where a contextual motion embedding (CME) is learned to encapsulate past motion states and the current environmental context, allowing it to produce rich, informative motion priors that enhance waypoint prediction accuracy and local motion tokens prediction. 
Instead of directly predicting the 6-DoF transformations, we employ a diffusion-based generative approach to sample realistic waypoints. Specifically, the model predicts the denoised sample at each step rather than predicting the noise added to the sample. This approach leverages a noise-to-signal refinement process, allowing more robust and diverse motion synthesis while preserving trajectory smoothness. The waypoint generation is structured hierarchically, where coarse-grained global waypoints are initially estimated before being refined into fine-grained local positions and orientations. This multi-resolution approach ensures robust motion trajectories and reduces high-frequency artifacts.

\subsection{Full Motion Sequence Generation}
The full motion sequence follows a hierarchical generation.
Once sparse waypoints are obtained and conditioned on text descriptions, scene voxel, and object point cloud, the same Transformer encoder backbone generates the full motion by refining and densifying the motion representation. The CME encoded from previous information will serve as prior planning for the model to generate more robust motion sequences. At this phase, the local motion tokenizer will encode the motion token at each time step to an integer index. The global human\&object 6DoF will be reshaped and stacked to be adapted with the temporal segment of the motion token. The 6-DoF representation will be passed by the same linear projector, and the motion token will be translated to continuous embedding with the same embedding layer used in the generation of the waypoints.

This two-stage process ensures that generated human-object motions are both physically plausible and semantically meaningful.
UniHM consists of two primary components: a self-supervised LFQ-VAE and a supervised autoregressive generative module. For the LFQ-VAE, we adhere to the following formulation for the total loss function:

\begin{equation}
\mathcal{L} = \lambda_{\text{CE}} \mathcal{L}_{\text{CE}} + \lambda_{\text{6DoF}} \mathcal{L}_{\text{6DoF}}
\end{equation}

To adapt the motion within the scene, our model must predict both the local motion token and the global 6-DoF for the human body and objects. We employ the cross-entropy loss $\mathcal{L}_{\text{CE}}$ for predicting motion tokens, defined as:

\begin{equation}
\mathcal{L}_{\text{CE}} = - \sum_{t} \sum_{k} p_t(k) \log \hat{p}_t(k)
\end{equation}

Here, $\hat{p}_t(k)$ represents the ground-truth token probability, while $p_t(k)$ is the predicted probability distribution at each timestep.

Conversely, the 6-DoF denoising loss employs a noise-to-sample refinement strategy:

\begin{equation*}
\mathcal{L}_{\text{6DoF}} = \mathbb{E}_{(w_0, \epsilon) \sim q} \left[ | f_{\theta}(w_t, t) - w_0 |^2 \right]
\end{equation*}

In this context, $x_0$ is the clean 6-DoF motion state, $x_t$ is the noisy sample, and the denoising network is parameterized by $\theta$. Unlike traditional diffusion models, UniHM directly predicts the clean sample instead of the noise. In practice, we found that a small number of denoise steps would be sufficient to get satisfactory results. Hence we set the denoise step to 20 using DDIM \cite{ddim} algorithms. To enhance controllability while maintaining generation diversity, like other defusion-based work, we apply classifier-free guidance (CFG) \cite{cfg} during the denoising process. Specifically, the final predicted sample is obtained by interpolating between the unconditioned and conditioned outputs. We applied classifier-free guidance by randomly replacing text conditions with null strings while training and calculating a combination of both conditioned and unconditioned generation results.
\begin{equation*}
    w_t^{\text{guided}} = w_t^{\text{uncond}} + s \cdot \left( w_t^{\text{cond}} - w_t^{\text{uncond}} \right),
\end{equation*}

where $s$ is the guidance scale that controls the strength of the conditioning signal. We do not replace the scene or the object for CFG since their information is relatively physically important and CFG may damage the fidelity of the physical behaviors.

In addition to the cross-entropy and 6-DoF denoising losses, we incorporate a geometry loss to enable the denoiser to learn more realistic Human-Object interactions. This is achieved by forcing the sampled object positions to have similar pairwise distances between the ground truth human bodies and ground truth object positions. 
It is worth noticing that the autoregressive model predicts the motion token logits instead of explicitly predicting the global motion. Directly calculating the pairwise distance between the ground truth human bodies and predicted object position, i.e. $\text{dist}(\hat{O}_i, J_i)$ may cause the overfitting and rigid object motion pattern. Therefore, we use a simple yet effective approach by decoding the same partially randomly replaced motion token and calculating the corresponding pairwise distance with the ground truth contact pattern, i.e. $\text{dist}(O_i, \hat{J}_i)$. Such distance

\begin{equation*}
\mathcal{L}_{\text{contact}} = \frac{1}{mn} \sum_{i=1}^{n}\sum_{j=1}^{m}||\text{dist}(O_i, J_j) - \text{dist}(\hat{O}_i, \hat{J}_j)||
\end{equation*}

Here, $m$ is the number of predefined points on the human mesh, and $n$ is the number of sampled points from object surfaces. $O_i/\hat{O}_i$ denotes the $i$-th sampled point on the ground truth/estimated objects, and $J_j/\hat{J}_j$ represents the $j$-th predefined point on the human mesh with partially replaced/predicted human motions.

\begin{table*}[htp]
\centering
\renewcommand{\arraystretch}{1.}
\caption{Comparison results of Text-to-HOI, and Text-to-HOI without Scene}
\resizebox{\textwidth}{!}{  
\begin{tabular}{lcccccccc}
\hline
\multicolumn{1}{c}{\multirow{2}{*}{Model}} & \multicolumn{1}{c}{\multirow{2}{*}{FID $\downarrow$}} & \multicolumn{3}{c}{R-precision $\uparrow$}                                              & \multicolumn{1}{c}{\multirow{2}{*}{MMD $\downarrow$}} & \multicolumn{1}{c}{\multirow{2}{*}{Diversity $\uparrow$}} & \multicolumn{1}{c}{\multirow{2}{*}{Collision $\downarrow$}} 
& \multicolumn{1}{c}{\multirow{2}{*}{Contact $\uparrow$}}\\

\multicolumn{1}{c}{}                        & \multicolumn{1}{c}{}                     & \multicolumn{1}{c}{top-1}                & \multicolumn{1}{c}{top-2} & \multicolumn{1}{c}{top-3} & \multicolumn{1}{c}{}     & \multicolumn{1}{c}{}   & \multicolumn{1}{c}{}             & \multicolumn{1}{c}{}                       \\
\hline
\multicolumn{9}{c}{\textbf{Text-to-Motion}} \\
\hline
Ground Truth & -  & 0.525  &  0.707   &    0.794       &  1.557   & 5.162 & - & - \\ 
\hline
MDM \cite{mdm} & 0.743 & 0.370 & 0.558 &  0.659 & 2.377 & 4.899 & - & - \\
MLD \cite{mld} & 0.412 & 0.401 & 0.582 &  0.680 & 2.134 & 4.920 & - & - \\
T2M-GPT \cite{t2mgpt} & \textbf{0.248} & {0.453} & {0.616} &  {0.712} & {1.873} & 5.038 & - & - \\
UniHM (ours)  & 0.302& \textbf{0.459} & \textbf{0.621} &  \textbf{0.729} & \textbf{1.785} & \textbf{5.124 } & - & - \\

\hline
\multicolumn{9}{c}{\textbf{Text-to-HOI}} \\
\hline
Ground Truth   &   -  & 0.579  &  0.796   &    0.890       &    5.515   & 5.521 & 0.002 & 1.000 \\ 
\hline
MDM & 1.392& 0.462 & 0.603 &  0.706 & 5.967 & 5.379 & 0.206 & 0.589\\
ARDHOI \cite{ardhoi} & {0.826}& {0.528} & {0.739} &  {0.843} & {5.774} &
\textbf{6.125} & {0.080} & 0.780\\
CHOIS \cite{chois} & 0.683& \textbf{0.553} &  {0.755} &   {0.860} & {5.642} &
5.973 & \textbf{0.036} & \textbf{0.874} \\
UniHM (ours) & \textbf{0.582}& {0.550} & \textbf{0.758} &  \textbf{0.862} & \textbf{5.639} & {6.034 } & {0.044} & {0.847}\\
\hline
\end{tabular}
} 
\label{tab:no_scene_result}
\end{table*}

\begin{table*}[htp]
\centering
\renewcommand{\arraystretch}{1.}
\caption{Comparison results of Text-to-Motion, Text-to-Motion in Scene. We add the scene voxel encoded by our ViT-VAE encoder to the variant marked with * as conditions}
\resizebox{\textwidth}{!}{  
\begin{tabular}{lccccccccc}
\hline
\multicolumn{1}{c}{\multirow{2}{*}{Model}} & \multicolumn{1}{c}{\multirow{2}{*}{FID $\downarrow$}} & \multicolumn{3}{c}{R-precision $\uparrow$}                                              & \multicolumn{1}{c}{\multirow{2}{*}{MMD $\downarrow$}} & \multicolumn{1}{c}{\multirow{2}{*}{ Diversity $\uparrow$}} & \multicolumn{1}{c}{\multirow{2}{*}{Collision\_o $\downarrow$}} 
& \multicolumn{1}{c}{\multirow{2}{*}{ Contact $\uparrow$}} & \multicolumn{1}{c}{\multirow{2}{*}{ Collision\_s $\downarrow$}} \\

\multicolumn{1}{c}{}                        & \multicolumn{1}{c}{}                     & \multicolumn{1}{c}{top-1}                & \multicolumn{1}{c}{top-2} & \multicolumn{1}{c}{top-3} & \multicolumn{1}{c}{}     & \multicolumn{1}{c}{}   & \multicolumn{1}{c}{}             & \multicolumn{1}{c}{}     & \multicolumn{1}{c}{}                  \\
\hline
\multicolumn{10}{c}{\textbf{Text-to-Motion in Scene}} \\
\hline
Ground Truth   &   -  & 0.462  &  0.646   &    0.741       &  2.140   & 5.370 & - & - & 0.000 \\ 
\hline
MDM   $*$ & 1.684 & 0.343 & 0.533 &  0.628 & 2.632 & 5.290 & - & - & 0.261 \\
MLD   $*$ & 1.429 & 0.368 & 0.550 &  0.635 & 2.602 & 5.134 & - & -  & 0.254 \\
T2M-GPT  $*$ & 0.683 & \textbf{0.402} & \textbf{0.582} &  \textbf{0.678} & \textbf{2.343} & 5.244 & - & - & 0.212 \\
UniHM (ours)  & \textbf{0.655} & \textbf{0.402} & 0.580 &  {0.671} & 2.369 & \textbf{5.370} & - & - & \textbf{0.017} \\
\hline
\multicolumn{10}{c}{\textbf{Text-to-HOI in Scene}} \\
\hline
Ground Truth   &   -  & 0.520  &  0.746   &    0.833       &    5.873   & 5.521 & 0.002 & 1.000 & 0.000 \\ 
\hline
MDM & 2.452 & 0.462 & 0.603 &  0.706 & 5.967 & 5.379 & 0.206 & 0.589 & 0.206\\
ARDHOI & {1.620}& {0.480} & {0.702} &  {0.792} &\textbf{5.774} &
\textbf{6.125} & {0.080} & 0.780 & 0.086 \\
CHOIS & 1.249 & \textbf{0.493} &  {0.705} &   \textbf{0.793} & {5.642} &
5.973 & \textbf{0.036} & \textbf{0.874} & 0.067\\
UniHM (ours) & \textbf{1.232} & {0.490} & \textbf{0.706} & {0.788} &  5.706 & 6.064 & 0.044 & 0.855 & \textbf{0.021} \\
\hline
\end{tabular}
}
\label{tab:scene_result}
\end{table*}

\section{Experiments}

\begin{figure*}[htp]
    \centering
    \includegraphics[width=1\linewidth]{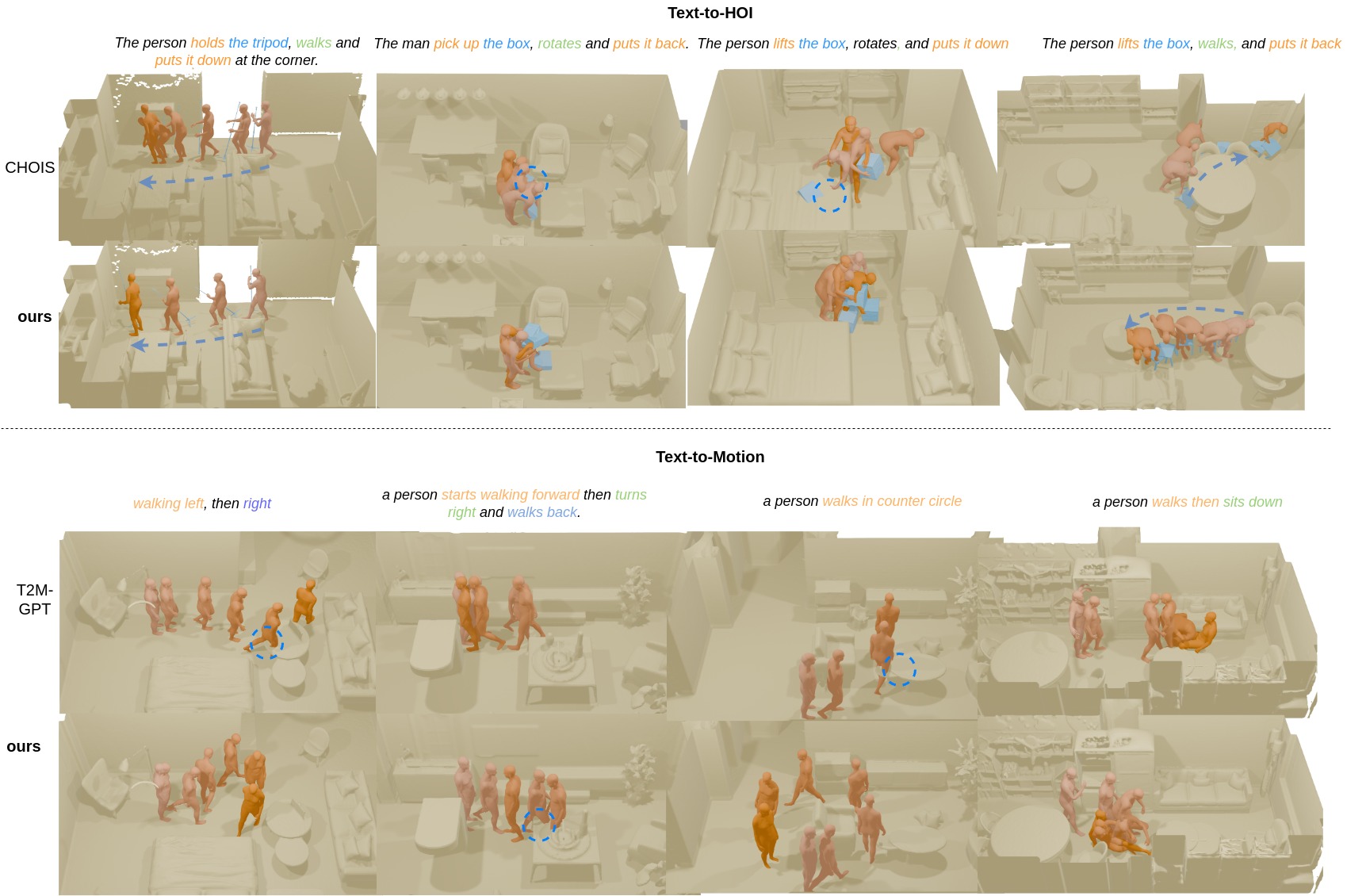}
    \caption{\textbf{Qualitative Comparison}. The first two rows are for Text-to-HOI, and the 3-rd \& the 4-th rows are for Text-to-Motion}
    \label{fig:cmp}
\end{figure*}

\subsection{Evaluation Metrics and Baselines}

We conduct extensive experiments across three primary tasks: Text-to-Motion, Text-to-Human-Object Interaction (Text-to-HOI), and Text-to-Human-Scene Interaction. These tasks are designed to assess our model’s ability to generate realistic motion sequences under different conditions. To fairly evaluate performance in both scene-free and scene-aware scenarios, we train our model on the entire UniHM dataset, ensuring that it learns representations applicable to both settings. For a fair comparison with prior works that do not incorporate scene information, we restrict their evaluation to scene-free motion sequences represented in a canonical form. In contrast, for evaluating models within a scene-aware context, we leverage our Scene ViT-VAE, which provides scene conditioning by learning meaningful spatial representations. The generated motion sequences are placed in the environment using the initial root position and orientation to ensure spatial consistency.

\par For Text-to-Motion, following prior works \cite{mdm, diversemotion, mld, t2mgpt}, we evaluate our model using several key metrics to measure the fidelity and diversity of the generated motions. These include:
\begin{itemize}
    \item \textbf{Frechet Inception Distance }(FID): This metric quantifies the distributional distance between generated motions and real motion sequences by comparing their high-level feature representations. A lower FID score indicates higher visual and structural realism. Recall Precision \item \textbf{R-Precision}: This metric measures the alignment between generated motions and textual descriptions by evaluating how closely the generated motion features match the corresponding text embeddings. A higher precision indicates better semantic correspondence. 
    \item \textbf{Multimodal Distance} (MMD): This computes the distance between the generated motion and text feature distributions, ensuring the generated content aligns with the given textual descriptions. 
    \item \textbf{Diversity Score}: This metric captures the variance in the generated motions, ensuring that the model does not collapse into generating repetitive outputs. A higher diversity score is desirable, as it indicates that the model produces a wider range of plausible motion sequences. 
\end{itemize}
To ensure comparability with previous research, we maintain the same evaluation protocol as prior studies but train our feature extractor using the SMPL motion representation instead of HumanML3D. This enables a more accurate comparison of motion synthesis quality.

\noindent \textbf{Text-to-HOI}. We follow similar evaluation protocols as used in Text-to-Motion but introduce additional metrics specifically designed to measure human-object interaction quality. These include:
\begin{itemize}
    \item 
\textbf{Human-Object Collision Score}: This metric assesses physical plausibility by measuring the extent to which the generated human motion collides with the object. It is computed using the signed distance field (SDF) of the object’s point cloud and predefined key points on the SMPL human mesh, as proposed in CHOIS \cite{chois}. Lower collision scores indicate more physically plausible interactions.
\item
\textbf{Contact Score}: This evaluates how well the human interacts with the object by counting the number of frames where the distance between the human body and the object’s point cloud is less than 5 cm. A higher contact score indicates that the model successfully captures realistic human-object interactions.
\end{itemize}

To validate the effectiveness of our approach, we compare it against previous state-of-the-art methods, including CHOIS \cite{chois} and ARDHOI \cite{ardiff}, which are specifically designed for modeling human-object interactions. Besides, for the metrics above, we also calculate the collision score of the scenes with both human bodies and objects for scene-aware settings.

\subsection{Implementation Details}

\noindent \textbf{LFQ-VAE} consists of 3 downsample layers each with a downsample rate of 2, compressing every 8 frames in a binary representation. The codebook has a total vocabulary of 8192 tokens the dimension of the base CNN is 128, and an expansion ratio of 1/2/4 for the 1st/2nd/3rd downsampling and upsampling. The coefficients of the objective function are $\lambda_{\text{recon}} = 1, \lambda_{\text{commit}} = 1\times 10^{-2}, \text{and} \lambda_{\text{entropy}} = 1\times 10^{-4}$

\noindent \textbf{UniHM} For our UniHM model, we adopt a causal Transformer encoder as the backbone architecture, which consists of 9 layers with 16 self-attention heads per layer. The model is configured with a hidden dimension of 512, a feedforward network (FFN) dimension of 2048, and a dropout rate of 0.1 to prevent overfitting. This architecture enables effective autoregressive motion generation while maintaining long-term coherence in synthesized motion sequences.

To enhance motion diversity and improve robustness in motion understanding, we introduce a token masking strategy during training. Specifically, we randomly select 20\% of the ground truth tokens and replace them with randomly sampled tokens. This forces the model to develop a stronger contextual understanding of motion patterns while ensuring that it does not overly rely on specific token dependencies. As a result, UniHM achieves better generalization across diverse motion categories, particularly in human-object interaction tasks.

\noindent \textbf{ViT-VAE} compresses the global information dimension of $2\times1024$. We utilize a ViT of 6 layers, 16 heads, a hidden dimension of 1024, and a feedforward dimension of 4096 for both the encoder and decoder of the VAE. For the decoder's output head, we equip it with an up-sampling network for reconstruction. Our ViT-VAE eventually achieved an accuracy of 66\% and an IOU of 60\%.

\begin{table*}[hbp]
\centering
\caption{Ablation Study on HSI dataset, waypoint, and denoiser for motion generation in scenes}
\label{tab:abl_scene}
\resizebox{\linewidth}{!}{
\begin{tabular}{lccccccccc}
\hline
\multicolumn{1}{c}{\multirow{2}{*}{Model}} & \multicolumn{1}{c}{\multirow{2}{*}{FID $\downarrow$}} & \multicolumn{3}{c}{R-precision $\uparrow$}                                              & \multicolumn{1}{c}{\multirow{2}{*}{MMD $\downarrow$}} & \multicolumn{1}{c}{\multirow{2}{*}{ Diversity $\uparrow$}} & \multicolumn{1}{c}{\multirow{2}{*}{ Collision\_s $\downarrow$}} 
& \multicolumn{1}{c}{\multirow{2}{*}{ Collision\_o  $\uparrow$}} & \multicolumn{1}{c}{\multirow{2}{*}{Contact $\uparrow$}}  \\

\multicolumn{1}{c}{}                        & \multicolumn{1}{c}{}                     & \multicolumn{1}{c}{top-1}                & \multicolumn{1}{c}{top-2} & \multicolumn{1}{c}{top-3} & \multicolumn{1}{c}{}     & \multicolumn{1}{c}{}   & \multicolumn{1}{c}{}             & \multicolumn{1}{c}{}                       \\
\hline
\multicolumn{10}{c}{\textbf{Text-to-Motion}} \\
\hline
Ground Truth   &   -  & 0.462  &  0.646   &    0.741       &  2.140  & 6.045 & - & - \\ 
\hline
UniHM (w/o. lingo) & \textbf{0.569}& \textbf{0.402} & {0.625} &  {0.724} & {1.840} & {6.153 } & 0.078 & - & - \\
UniHM (w/o. way point) & {0.684}& {0.389} & {0.613} &  {0.720} & {1.866} & {5.060 } & 0.152 & - & - \\
UniHM (w/o. VIT pre-trained) & {0.695}& {0.397} & {0.618} &  {0.722} & {1.856} & {5.065 } & 0.083 & - & - \\
UniHM (w/o. denoiser)  & {0.690}& {0.400} & {0.620} &  {0.726} & {1.830} & {4.964 } & 0.022 & - & - \\
UniHM (ours)  & {0.655} & \textbf{0.402} & 0.580 &  {0.671} & 2.369 & \textbf{5.370} & \textbf{0.017} & - & - \\
\hline
\multicolumn{10}{c}{\textbf{Text-to-HOI}} \\
\hline
Ground Truth   &   -  & 0.520  &  0.746   &    0.833       &    5.873   & 5.521 & 0.000 & 1.000 & 0.002 \\ 
\hline
UniHM (w/o. lingo) & {1.154}& {0.502} & {0.640} &  {0.825} & {5.948} & {5.973} & 0.088 & 0.102 & 0.695\\
UniHM (w/o. way point) & {1.056}& {0.507} & {0.646} &  {0.830} & {5.934} & {5.934} & 0.120 & 0.146 & 0.630\\
UniHM (w/o. VIT pre-trained) & {0.725}& {0.539} & {0.717} &  {0.843} & {1.856} & {5.065 } & 0.090 & - & - \\
UniHM (w/o. denoiser) & {0.642}& {0.545} & {0.746} &  \textbf{0.862} & {5.689} & {5.987 } & 0.025 & {0.056} & \textbf{0.853}\\
UniHM (ours) & \textbf{1.232} & {0.490} & \textbf{0.706} & {0.788} &  5.706 & 6.064 & \textbf{0.021} & 0.044 & 0.855  \\
\hline
\end{tabular}
}
\end{table*}

\subsection{Quantative Results}

We compared our result with MDM \cite{mdm}, MLD \cite{mld}, and T2M-GPT \cite{t2mgpt} for comparison on Text-to-Motion with/without scenes; and compared with MDM, ARDHOI \cite{ardhoi}, and CHOIS\cite{chois}. We modify the MDM in the scene-aware task with our scene ViT encoder to enable the MDM scene awareness. It should be noticed that CHOIS utilizes the ground truth waypoint to guide a realistic contact while inference, for fairness for generation from scratch, we remove such conditions.

The results in Table \ref{tab:no_scene_result} suggest that our approach outperforms T2M-GPT for the Text-to-Motion task in most metrics but for FID. Such a result is probably caused by the generation of the waypoint and 6-DoF that share the model capacity for predicting the token. The results for Text-to-HOI show that our approach achieves better performance than CHOIS in FID, R-precisions, and Multimodal Distance. Although our method has no significant contact sampling such as classifier guidance, we still achieved comparable contact and collision performance. 

For the scene-aware task, we illustrate our results in Table \ref{tab:scene_result}. The results suggest that our method achieves comparable results with ARDHOI and CHOIS and significantly decreases the collision with the scene. 

\subsection{Qualatative Comparison}

We visualize the result for both Text-to-Motion and Text-to-HOI in scenes. For the Text-to-Motion task, we compared the result with T2M-GPT and for the Text-to-HOI task, we compared the result with CHOIS. The results in Fig. \ref{fig:cmp} suggest that our model can better generate reasonable motion in the scene. For the Text-to-Motion task, our approach which generates a reasonable waypoint sequence in the scenes effectively reduces the collision with the scene. 

For the Text-to-HOI task, with the participation of the predicted waypoints and replaced motion tokens, our model can generate satisfactory and robust human-object interaction. 

\begin{figure}[htp]
    \centering
    \includegraphics[width=0.8\linewidth]{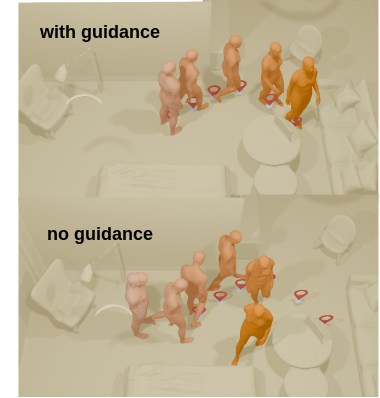}
    \caption{\textbf{Motions generated in the scene with/without waypoint guidance}}
    \label{fig:waypoint_guide}
\end{figure}

\begin{figure}
    \centering
        \centering
        \includegraphics[width=0.8\linewidth]{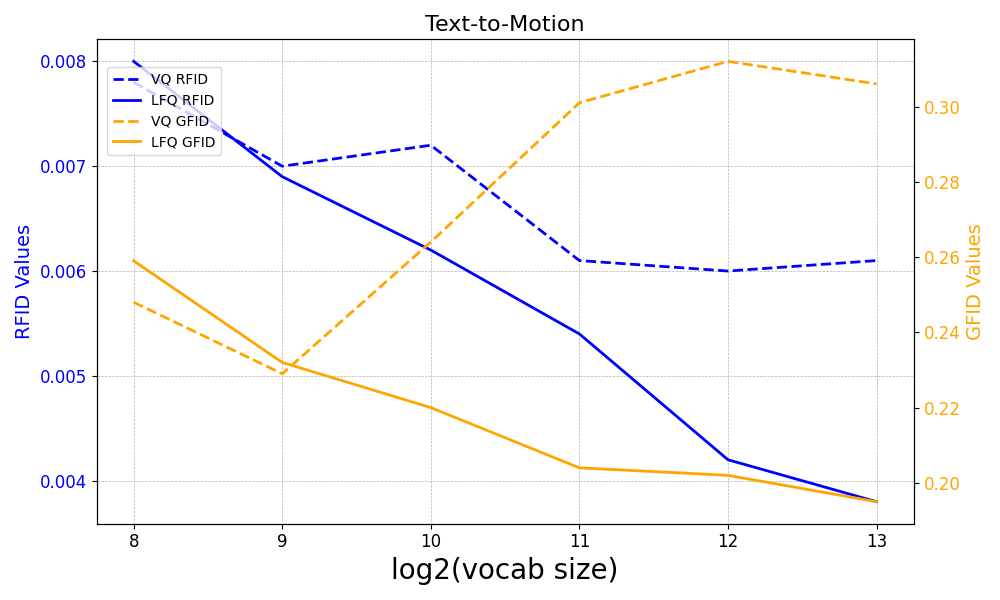}
        \label{fig:rfid}
    
        \centering
        \includegraphics[width=0.8\linewidth]{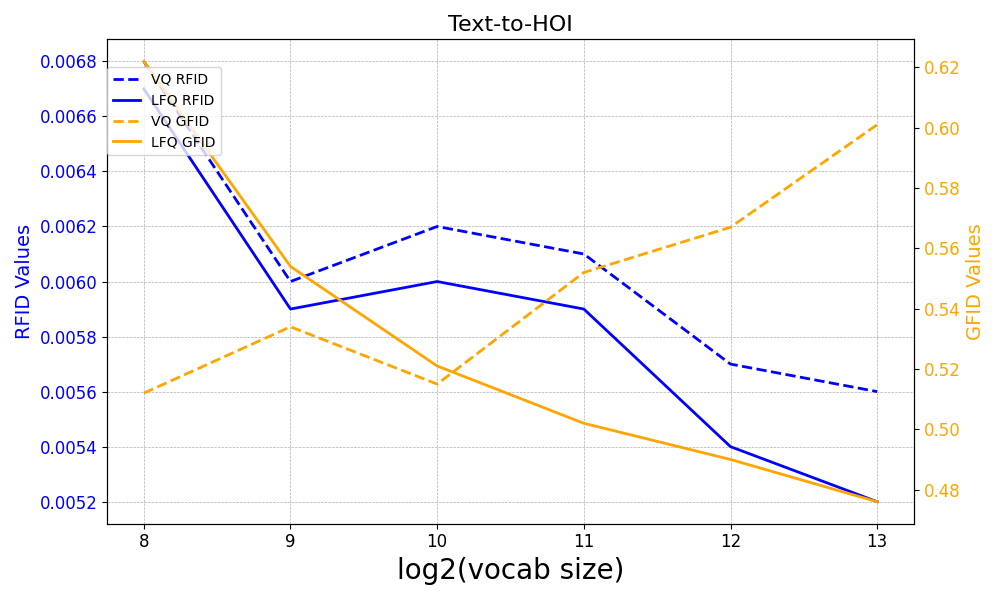}
        \label{fig:rfid_hoi}
    
    \caption{\textbf{Comparison of rFID and gFID for VQ-VAE and LFQ-VAE across different tasks.}}
    \label{fig:rfid_all}
\end{figure}

\begin{table}[htp]
\centering
\caption{Performance of waypoint guidance in scenes}
\label{tab:wpguide}
\begin{tabular}{lcccc}
\hline
{Model} & {FID $\downarrow$} & {Collision\_s $\downarrow$}
& {Collision\_o  $\uparrow$} & {Contact $\uparrow$}  \\
\hline
\multicolumn{5}{c}{\textbf{Text-to-Motion}} \\
\hline
Ground Truth   &   -  & - & - & - \\ 
\hline
UniHM & {0.655} & 0.017 & - & - \\
UniHM (guided) & \textbf{0.310} & \textbf{0.010} & - & - \\
\hline
\multicolumn{5}{c}{\textbf{Text-to-HOI}} \\
\hline
Ground Truth   &   -  & 0.000 & 1.000 & 0.002 \\ 
\hline
UniHM & {1.232} & 0.021 & 0.044 & 0.855 \\
UniHM (guided) & \textbf{0.826} & \textbf{0.015} & \textbf{0.028} & \textbf{0.945}\\

\hline
\end{tabular}
\end{table}

\begin{table*}[htp]
\centering
\renewcommand{\arraystretch}{1.}
\caption{Ablation Study on HSI dataset, waypoint, and denoiser for Text-to-Motion and Text-to-HOI without scene}
\label{tab:abl}
\resizebox{\linewidth}{!}{  
\begin{tabular}{lcccccccc}
\hline
\multicolumn{1}{c}{\multirow{2}{*}{Model}} & \multicolumn{1}{c}{\multirow{2}{*}{FID $\downarrow$}} & \multicolumn{3}{c}{R-precision $\uparrow$}                                              & \multicolumn{1}{c}{\multirow{2}{*}{MMD $\downarrow$}} & \multicolumn{1}{c}{\multirow{2}{*}{ Diversity $\uparrow$}} & \multicolumn{1}{c}{\multirow{2}{*}{ Collision $\downarrow$}} 
& \multicolumn{1}{c}{\multirow{2}{*}{ Contact $\uparrow$}}\\

\multicolumn{1}{c}{}                        & \multicolumn{1}{c}{}                     & \multicolumn{1}{c}{top-1}                & \multicolumn{1}{c}{top-2} & \multicolumn{1}{c}{top-3} & \multicolumn{1}{c}{}     & \multicolumn{1}{c}{}   & \multicolumn{1}{c}{}             & \multicolumn{1}{c}{}                       \\
\hline
\multicolumn{9}{c}{\textbf{Text-to-Motion}} \\
\hline
Ground Truth   &   -  & 0.462  &  0.646   &    0.741       &  1.557   & 6.045 & - & - \\ 
\hline
UniHM (w/o. lingo) & \textbf{0.300}& {0.452} & {0.625} &  {0.724} & {1.840} & {6.153 } & - & - \\
UniHM (w/o. way point) & {0.389}& {0.428} & {0.613} &  {0.720} & {1.866} & {5.060 } & - & - \\
UniHM (w/o. denoiser)  & {0.367}& {0.444} & {0.620} &  {0.726} & {1.830} & {4.964 } & - & - \\
UniHM (w. denoiser)  & {0.302}& \textbf{0.459} & \textbf{0.621} &  \textbf{0.729} & \textbf{1.785} & \textbf{5.124 } & - & - \\
\hline

\hline
\multicolumn{9}{c}{\textbf{Text-to-HOI}} \\
\hline
Ground Truth   &   -  & 0.579  &  0.796   &    0.890       &    5.515   & 5.521 & 0.002 & 1.000 \\ 
\hline
UniHM (w/o. lingo) & {1.154}& {0.502} & {0.640} &  {0.835} & {5.948} & {5.973} & 0.102 & 0.695\\
UniHM (w/o. way point) & {1.056}& {0.507} & {0.646} &  {0.840} & {5.934} & {5.934} & 0.146 & 0.630\\
UniHM (w/o. denoiser) & {0.642}& {0.545} & {0.746} &  \textbf{0.862} & {5.689} & {5.987 } & {0.056} & \textbf{0.853}\\
UniHM (w. denoiser) & \textbf{0.582}& \textbf{0.550} & \textbf{0.758} &  \textbf{0.862} & \textbf{5.639} & \textbf{6.034 } & \textbf{0.044} & {0.847}\\
\hline
\end{tabular}
} 
\end{table*}

\subsection{Waypoint Guidance}

A key advantage of our proposed method lies in its ability to integrate waypoint guidance, which plays a crucial role in significantly enhancing the fidelity and coherence of generated motions, particularly within the context of autoregressive hierarchical generation. Unlike conventional motion generation approaches that rely solely on learned latent representations, our method introduces intermediate waypoint constraints, ensuring that the generated motion remains physically plausible and temporally consistent. As illustrated in Fig. \ref{fig:waypoint_guide}, our model can effectively utilize predefined 6-DoF waypoints to guide the motion synthesis process, leading to more structured and realistic movement trajectories.

While this waypoint-based guidance does not inherently guarantee exact 6-DoF control—especially when compared to diffusion models with classifier guidance \cite{gmd}, which employ gradient-based optimization mechanisms for precise control—it still brings substantial improvements in motion quality. The classifier-free guidance approach in diffusion models explicitly refines the generated samples based on gradients derived from an external classifier. In contrast, our approach ensures smoother transitions and more coherent motion sequences by embedding waypoints directly into the generation process, without requiring iterative refinement steps.

Furthermore, by incorporating ground truth 6-DoF waypoints prior to generating full motion sequences, our method effectively constrains the trajectory to adhere to a more meaningful and natural motion path. This results in a marked increase in accuracy and realism, as reflected in key evaluation metrics. As demonstrated in Table \ref{tab:wpguide}, our method achieves notable improvements in FID, collision avoidance, and contact scores, reaffirming the effectiveness of waypoint guidance in refining generated motion sequences. These improvements highlight the robustness of our approach, making it particularly suitable for applications requiring precise motion synthesis, such as human-object interactions and scene-aware motion generation.

\subsection{LFQ-VAE vs. VQ-VAE}

To evaluate the efficacy of LFQ-VAE against VQ-VAE, we compared their generation FID (gFID) and reconstruction FID (rFID). Additionally, we applied both models to the common Text-to-Motion tasks without scene context to minimize variability and ensure a fair comparison. The results in Fig. \ref{fig:rfid_all} suggest that with the increase of the vocabulary size, both the rFID for VQ-VAE and LFQ-VAE improved but the gFID for VQ-VAE increased while the LFQ-VAE continues improving.

\subsection{Ablation Study}

We conduct the ablation study on each component and design methods including waypoint generation, denoiser functionality, and lingo dataset efficacy. The first line in Table \ref{tab:abl} suggests that with the Lingo Dataset,  performance has a significant increase in Multimodal Distance and R-precision, but a slight decrease in FID. This might be due to the intrinsic difference in the motion distribution between the HumanML3D dataset and the Lingo dataset. Noticing the fact that HumanML3D focuses on macro motion with clear semantic information, and Lingo Dataset focuses more on indoor motion, we can explain the overall increased performance evaluated on the OMOMO dataset, which has a more similar motion origin with Lingo. What's more, the results in the second line suggest that without waypoint generation, the FID increases marginally than the group with hierarchical waypoint generation.

\section{Conclusion}

In this work, we introduced UniHM, a universal human motion generation model that integrates text, waypoints, and scene contexts to synthesize motion in complex environments. To address the limitations of motion tokenization, we proposed a mixed-motion representation combining continuous 6DoF motion with discrete local motion tokens. Additionally, we developed LFQ-VAE, an improved quantization method that enhances motion representation fidelity and generation quality. To support scene-aware motion synthesis, we augmented existing datasets with HumanML3D annotations, enriching scene-conditioned motion learning. Our approach advances scene-based human motion synthesis, bridging the gap between Text-to-Motion and realistic scene interactions, with promising results on HumanML3D and OMOMO datasets.

\bibliographystyle{IEEEtran}
\bibliography{unihm}

\begin{IEEEbiography}[{\includegraphics[width=1in,height=1.25in,keepaspectratio]{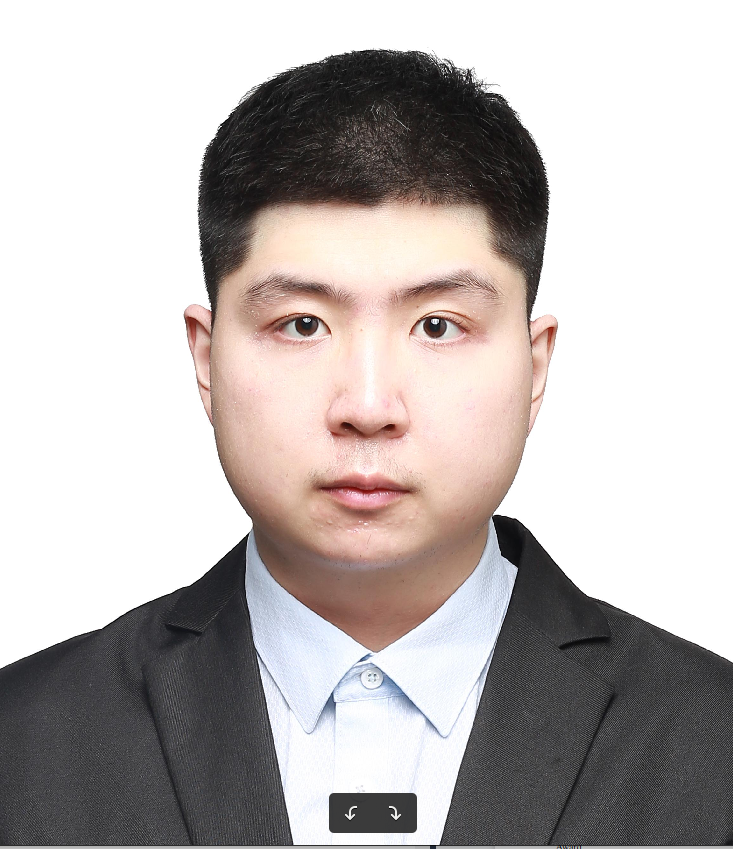}}]{Zichen Geng} received the B.E degree in Shanghai Jiao Tong University and received the M.S degree in the University of Western Australia. Now he is pursuing the Ph.D. degree in Computer Science at the University of Western Australia, supervised by Prof. Ajmal Mian and A/Prof. Wei Liu. His research interests include human motion and generative models.
\end{IEEEbiography}


\begin{IEEEbiography}[{\includegraphics[width=1in,height=1.25in,clip,keepaspectratio]{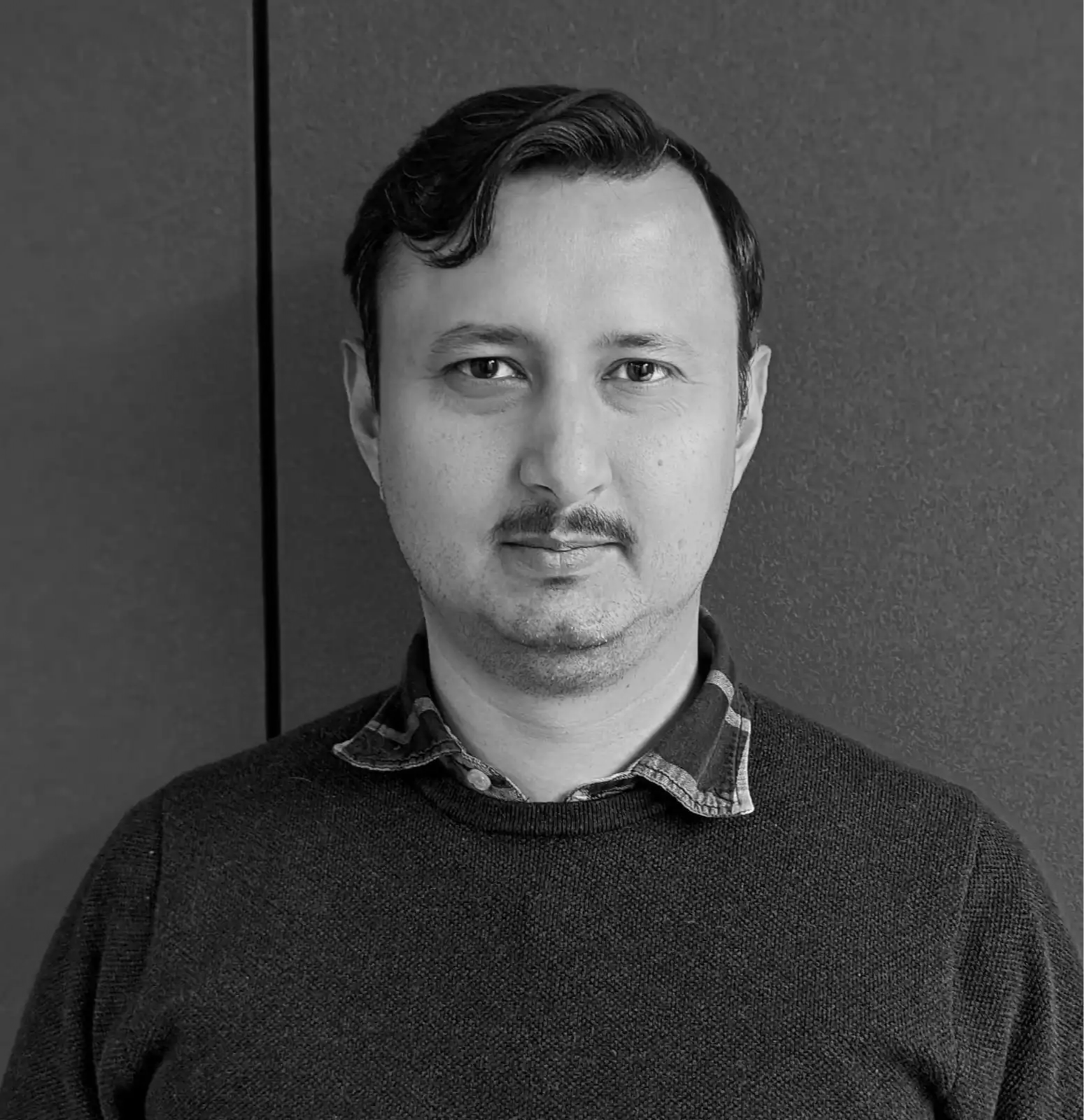}}]{Zeeshan Hayder} is currently a Research Scientist in Data61, CSIRO, Australia. He received a Ph.D. in Computer Engineering from The Australian National University in 2018. He has extensive industry experience, specialising in machine learning, computer vision, and artificial intelligence. His research interests include vision-language modeling, learning with limited labels, and self-supervised learning.
\end{IEEEbiography}

\begin{IEEEbiography}
[{\includegraphics[width=1in,height=1.25in,keepaspectratio]{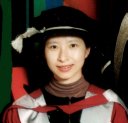}}]{Wei Liu} is an Associate Professor in the University of Western Australia. She received her PhD from the University of Newcastle, Australia in 2003. She is now a full-time teaching \& research academic in the Department of Computer Science and Software Engineering at the University of Western Australia. She leads the UWA Centre for Natural and Technical Language Processing, with research effort focusing on knowledge discovery from natural language text (NLP), generative AI, LLM-based semantic technologies, deep learning methods for knowledge graph construction and analysis, as well as sequential data mining and forecasting. Her recent work includes training machine learning models from fusing multi-modality heterogeneous data, and neural-symbolic computation with human-in-the-loop. Her industry-related research projects include knowledge graph refinement for geological survey reports, technical language processing on maintenance work orders, incident/safety log analysis and visualization, short-term traffic prediction, and clinical data integration and analysis in ophthalmology.
\end{IEEEbiography}

\begin{IEEEbiography}[{\includegraphics[width=1in,height=1.25in,keepaspectratio]{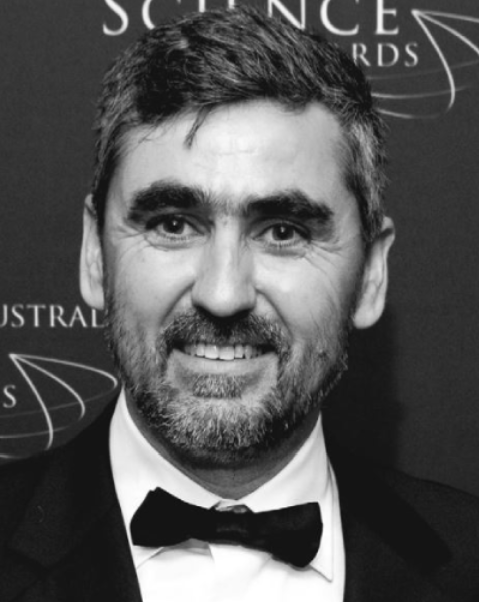}}]{Ajmal Mian} (Senior Member, IEEE) is currently a
Professor of computer science with The University
of Western Australia, Perth, WA, Australia. He has
received several major research grants from the
Australian Research Council (ARC), the National
Health and Medical Research Council of Australia,
and the U.S. Department of Defense. His research
interests include computer vision, machine learning,
and artificial intelligence.
Prof. Mian is an Australian Research Council
Future Fellow, an International Association for
Pattern Recognition (IAPR) Fellow, an ACM Distinguished Speaker, and
the President of the Australian Pattern Recognition Society. He has received
several awards, including the West Australian Early Career Scientist of the
Year Award, the Vice-Chancellors Mid-Career Research Award, the Outstanding Young Investigator Award, the IAPR Best Scientific Paper Award, the
EH Thompson Award, and the Excellence in Research Supervision Award.
He served as the General Chair for Asian Conference on Computer Vision
(ACCV) 2018 and International Conference on Digital Image Computing:
Techniques and Applications (DICTA) 2019 and the Area Chair for ACM
International Conference on Multimedia (ACM MM) 2020, IEEE Conference
on Computer Vision and Pattern Recognition (CVPR) 2022, and European
Conference on Computer Vision (ECCV) 2022. He now serves as a Senior
Editor for IEEE TRANSACTIONS ON NEURAL NETWORKS AND LEARNING
SYSTEMS (TNNLS) and an Associate Editor for IEEE TRANSACTIONS ON
IMAGE PROCESSING (TIP) and the Pattern Recognition (PR) journal.
\end{IEEEbiography}


\vfill

\end{document}